# A study on refractive index sensors based on optical micro-ring resonators

Georgios N. Tsigaridas[1]

Department of Physics, School of Applied Mathematical and Physical Sciences, National Technical University of Athens, Zografou Campus, GR-15780 Zografou, Athens, Greece

**Abstract**

In this work the behavior of optical micro-ring resonators, especially when functioning as refractive index sensors, is studied in detail. Two configurations are considered, namely a linear waveguide coupled to a circular one and two linear waveguides coupled to each other through a circular one. The optimum coupling conditions are derived and it is shown that in both cases the condition for the resonant wavelength, i.e. the wavelength at which the transmission spectrum exhibits a dip (peak), is the same and depends only on the geometrical characteristics of the circular waveguide and the effective refractive index of the propagating mode. The latter, as well as the corresponding mode profile, can be easily calculated through numerical analysis. The sensitivity of the sensor is defined based on the dependence of the effective refractive index on the refractive index of the environment. Using a result of waveguide perturbation theory, the geometrical characteristics of the core of the circular waveguide that maximize the sensitivity of the system are determined. Both single and dual core configurations are considered. It is found that, when optimally designed, the sensor can detect relative refractive index changes of the order of $\Delta n/n \approx 10^{-4}$, assuming that the experimental setup can detect relative wavelength shifts of the order of $\Delta\lambda/\lambda \approx 3\times 10^{-5}$. Finally, the behavior of the system as bio-sensor is examined by considering that a thin layer of bio-material is attached on the surface of the waveguide core. It is found that, when optimally designed, the system can detect refractive index changes of the order of $\Delta n \approx 10^{-3}$ for a layer thickness of $t = 10\ nm$, and changes in the layer thickness of the order of $\Delta t \approx 0.24\ nm$, for a refractive index change of $\Delta n = 0.05$.

---

[1] E-mail: gtsig@mail.ntua.gr

## 1. Introduction and theoretical analysis

Optical ring resonators are interesting optical devices with a plethora of applications especially in optical switching [1-4], routing [4-7], and sensing [8-10]. An optical ring resonator usually consists of a straight waveguide coupled to a circular one, as shown in figure 1(a), or two straight waveguides coupled through a circular one, as shown in figure 1(b).

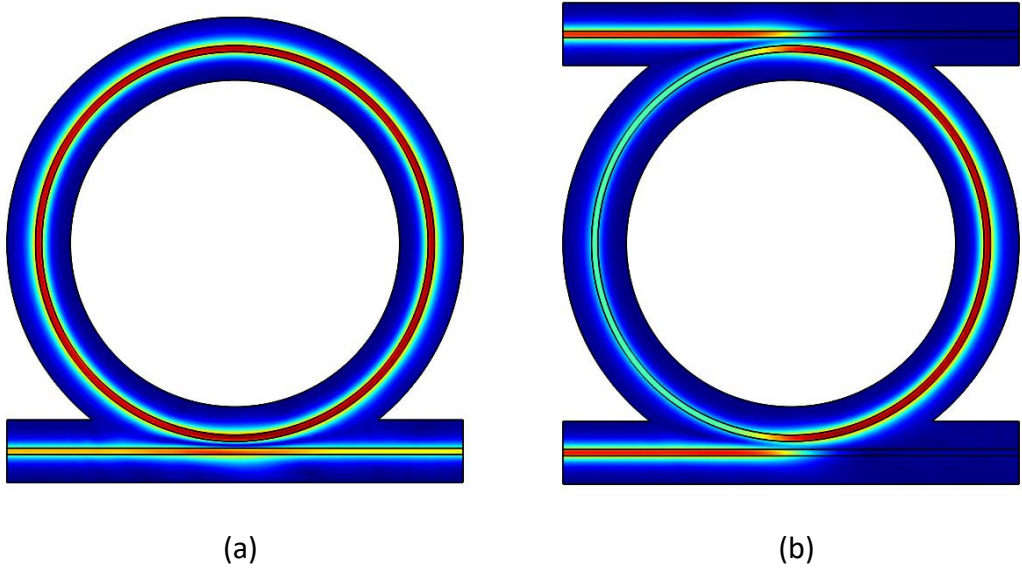

(a)  (b)

**Figure 1**: *Two basic configurations of an optical micro-ring resonator: (a) a linear waveguide is coupled to a circular one, and (b) two linear waveguides are coupled to each other through a circular one.*

In the first case the transmittance is given by the formula [11]

$$T_n = \frac{a^2 + r^2 - 2ar\cos\varphi}{1 + r^2 a^2 - 2ar\cos\varphi} \tag{1}$$

where $r$ is the self-coupling coefficient of light between the straight and circular waveguide, and $a$ is a loss parameter related to the power attenuation coefficient $\alpha$ through the equation $a^2 = \exp(-\alpha L)$, where $L$ is the length of the circular waveguide. Here, $\varphi$ is the single-pass phase shift, defined as

$$\varphi = \beta L = \frac{2\pi n_{eff}}{\lambda_0} L \tag{2}$$

where $\beta$ is the propagation constant and $n_{eff}$ the effective refractive index of the propagating mode in the circular waveguide, and $\lambda_0$ is the free space wavelength. Setting $\cos\varphi \to y$, equation (1) takes the form

$$T_n = \frac{a^2 + r^2 - 2ary}{1 + r^2 a^2 - 2ary} \tag{3}$$

The derivative of $T_n$ with respect to $y$ is

$$\frac{dT_n}{dy} = -\frac{2a(-1+a^2)r(-1+r^2)}{(1+a^2 r^2 - 2ary)^2} \tag{4}$$

It is obvious that $dT_n/dy \leq 0$ since both $a, r$ are less or equal to unity. Thus, the transmittance is minimized for $y = 1$, which implies that the condition for minimum transmittance as far as the phase shift is concerned is

$$\cos\varphi = 1 \text{ or } \varphi = 2m\pi \tag{5}$$

where m is an integer. Through equations (2), (5) follows that the wavelengths for minimum transmittance, resonant wavelengths, are given by the formula

$$\lambda_{0,res} = \frac{L}{m} n_{eff} = \frac{2\pi R}{m} n_{eff} \tag{6}$$

where $R$ is the radius of the circular waveguide. In this case the transmittance takes the form

$$T_{n,res} = \left(\frac{a-r}{1-ar}\right)^2 \tag{7}$$

which implies that for $a = r$, critical coupling, the transmittance is zero.

As far as the arrangement shown in fig. 1(b) is concerned, the transmittance through the lower and upper waveguides are given by the formulae

$$T_p = \frac{r_1^2 + a^2 r_2^2 - 2a r_1 r_2 y}{1 + a^2 r_1^2 r_2^2 - 2a r_1 r_2 y} \tag{8a}$$

and

$$T_d = \frac{a(1-r_1^2)(1-r_2^2)}{1 + a^2 r_1^2 r_2^2 - 2a r_1 r_2 y} \tag{8b}$$

respectively. Here $r_1, r_2$ are the coupling coefficients between the lower-circular and upper-circular waveguides respectively and $y = \cos\varphi$. Following a similar analysis it can be shown that the condition for minimum transmittance through the lower waveguide, and simultaneously maximum transmittance through the upper waveguide, is described again by equation (6). In this case equations (8a), (8b) take the form

$$T_{p,res} = \left(\frac{r_1 - ar_2}{1 - ar_1r_2}\right)^2 \tag{9a}$$

and

$$T_{d,res} = \frac{a(1-r_1^2)(1-r_2^2)}{(1-ar_1r_2)^2} \tag{9b}$$

Equation (9a) implies that for $r_1 = ar_2$, $T_p = 0$. In this case the transmittance through the upper waveguide takes the form

$$T_{d,res}\big|_{r_1=ar_2} = \frac{a(r_2^2 - 1)}{a^2 r_2^2 - 1} \tag{10}$$

It is obvious that for $a = 1$, $T_d = 1$. Thus, in the case of a symmetrical optical ring resonator, shown in figure 1(b), the light is perfectly coupled to the upper waveguide, when the wavelength is given by equation (6), the loss parameter is equal to unity $(a = 1)$, and the coupling coefficients are equal to each other $(r_1 = r_2)$.

As far as the use of micro-ring resonators as refractive index sensors is concerned, the most important result of the above analysis is that, in both configurations, the resonant wavelength is given by equation (6), which practically implies that it is proportional to the effective refractive index of the propagating mode in the circular waveguide. However, the value of $n_{eff}$ depends on the refractive index of the environment. Therefore, the system can function as a refractive index sensor. In more detail, supposing that the effective refractive index changes by $\Delta n_{eff}$ due to a change of the refractive index of the environment by $\Delta n$, equation (6) takes the form

$$\lambda_{0,res} + \Delta\lambda_{0,res} = \frac{2\pi R}{m}\left(n_{eff} + \Delta n_{eff}\right) \tag{11}$$

where $\Delta\lambda_{0,res}$ is the corresponding change in the resonant wavelength. Dividing equations (11), (6) by parts results that

$$\frac{n_{eff} + \Delta n_{eff}}{n_{eff}} = \frac{\lambda_{0,res} + \Delta\lambda_{0,res}}{\lambda_{0,res}} \quad \text{or} \quad \frac{\Delta n_{eff}}{n_{eff}} = \frac{\Delta\lambda_{0,res}}{\lambda_{0,res}} \tag{12}$$

Consequently, the relative change in the resonant wavelength is equal to the relative change in the effective refractive index.

According to the above analysis, the most important parameter regarding the performance of an optical ring resonator when used as a refractive index sensor is the dependence of the effective refractive index of the propagating mode on the refractive index of the environment. Mathematically, this is described by the quantity

$$S = \frac{1}{n_{eff}} \frac{dn_{eff}}{dn} \quad \text{or} \quad S = \frac{1}{n_{eff}} \frac{\Delta n_{eff}}{\Delta n} \tag{13}$$

where in the second form of equation (13) it is supposed that the refractive index change of the environment is sufficiently small, so that the dependence of $\Delta n_{eff}$ on $\Delta n$ is linear. Using the parameter $S$, equation (12) takes the form

$$\frac{\Delta \lambda_{0,res}}{\lambda_{0,res}} = S \Delta n \tag{14}$$

From equation (14) it is obvious that $S$ can be regarded as a measure of the sensitivity of the system when used as a refractive index sensor. In practice, the parameter $S$ can be numerically calculated quite easily as described in the following section.

## 2. Numerical analysis

The profile and effective refractive indexes of the propagating modes in the circular waveguide are calculated using finite element analysis [12, 13]. The simulations were assisted by the Comsol Multiphysics software package [14]. The analysis is made on a two dimensional cross-section of the waveguide, assuming that its profile and properties remain constant throughout its length. First, the electromagnetic field is written in the form

$$E(x,y,z) = A(x,y)\exp(-\alpha_0 z)\exp(-i\beta z) = A(x,y)\exp(-i\gamma z) \tag{15a}$$

$$H(x,y,z) = B(x,y)\exp(-\alpha_0 z)\exp(-i\beta z) = B(x,y)\exp(-i\gamma z) \tag{15b}$$

where $E(x,y,z)$, $H(x,y,z)$ are the electric and magnetic field components of the propagating wave respectively. The two-dimensional functions $A(x,y)$, $B(x,y)$ describe the mode profile, assuming that the electromagnetic wave is propagating in the $+z$ direction. Here $\alpha_0$ is the linear loss coefficient and $\beta$ the propagation constant, related to the effective refractive index through the formula

$$\beta = k_0 n_{eff} \tag{16}$$

where $k_0 = 2\pi/\lambda_0$ is the free space wavenumber. The parameter $\gamma$ corresponds to the generalized propagation constant, defined as $\gamma = \alpha_0 - i\beta$.

Writing the Maxwell equations in matrix form, and eliminating the longitudinal field components, an eigenvalue problem of the form [13]

$$\boldsymbol{\Omega}^2 \begin{bmatrix} \mathbf{A}_x \\ \mathbf{A}_y \end{bmatrix} = \left(\frac{\gamma}{k_0}\right)^2 \begin{bmatrix} \mathbf{A}_x \\ \mathbf{A}_y \end{bmatrix} \tag{17}$$

is obtained, where

$$\Omega^2 = \begin{bmatrix} \mathbf{D}^e_{x'}\varepsilon_{zz}^{-1}\mathbf{D}^h_{y'} & -\left(\mathbf{D}^e_{x'}\varepsilon_{zz}^{-1}\mathbf{D}^h_{x'} + \mu_{yy}\right) \\ \mathbf{D}^e_{y'}\varepsilon_{zz}^{-1}\mathbf{D}^h_{y'} + \mu_{xx} & \mathbf{D}^e_{y'}\varepsilon_{zz}^{-1}\mathbf{D}^h_{x'} \end{bmatrix} \begin{bmatrix} \mathbf{D}^h_{x'}\mu_{zz}^{-1}\mathbf{D}^e_{y'} & -\left(\mathbf{D}^h_{x'}\mu_{zz}^{-1}\mathbf{D}^e_{x'} + \varepsilon_{yy}\right) \\ \mathbf{D}^h_{y'}\mu_{zz}^{-1}\mathbf{D}^e_{y'} + \varepsilon_{xx} & \mathbf{D}^h_{y'}\mu_{zz}^{-1}\mathbf{D}^e_{x'} \end{bmatrix}$$

(18)

Here $\mathbf{D}^e_{x'(y')}, \mathbf{D}^h_{x'(y')}$ are matrix differential operators with respect to $x'(y')$ for the electric and magnetic field respectively, and $\varepsilon_{xx}, \varepsilon_{yy}, \varepsilon_{zz}, \mu_{xx}, \mu_{yy}, \mu_{zz}$ are matrices containing the diagonal elements of the relative permittivity and permeability tensors at each position. The vector matrix $\begin{bmatrix} \mathbf{A}_x \\ \mathbf{A}_y \end{bmatrix}$ corresponds to the transverse distribution of the electric field, $A(x,y)$, with $\mathbf{A}_x$, $\mathbf{A}_y$ containing – in single column format – the x- and y- components of the electric field at each position. The coordinates $(x', y', z')$ are related to $(x, y, z)$ through the transformation $(x' \to k_0 x, y' \to k_0 y, z' \to k_0 z)$.

Thus, solving equation (17) for the eigenvalues $\lambda = (\gamma/k_0)^2$ and the eigenvectors $\begin{bmatrix} \mathbf{A}_x \\ \mathbf{A}_y \end{bmatrix}$, the generalized propagation constant – and consequently the effective refractive index – as well as the transverse electric field distribution (profile) for each mode is obtained.

It should be noted that in most applications of practical interest, single mode operation is preferable, because the existence of more than one propagating modes would complicate the behavior of the system. For example, in the case of multi-mode operation, equation (6) implies that more than one resonant wavelengths exist for each configuration, which would result to multiple dips (or peaks, if measured through the upper waveguide) in the transmittance spectrum, complicating both the theoretical and experimental analysis of the system. Therefore, in the simulations, special care is taken in order to ensure single mode operation.

### 3. Simulations, results and discussion regarding design optimization

According to a result of waveguide perturbation theory [15], the change in the effective refractive index of the propagating mode due to a refractive index change in some part of the waveguide is proportional to the fraction of the mode power in this specific part of the waveguide, defined as

$$\eta_p = \frac{\iint\limits_{S_p} |A(x,y)|^2 \, dxdy}{\iint\limits_{S_{tot}} |A(x,y)|^2 \, dxdy} \tag{18}$$

where $S_p$ is the area where the refractive index has changed and $S_{tot}$ is the total area of the waveguide. Further, according to equation (13) the sensitivity $S$ is proportional to $\Delta n_{eff}$ and consequently it will also be proportional to $\eta_p$. This practically means that the sensitivity increases as the overlapping of the propagating mode with its environment becomes more extensive.

Therefore in the simulations the core of the waveguide is considered to be in direct contact with the environment. Further the core dimensions are chosen to be small, close to the limit of waveguiding, in order to increase the mode spreading and maximize the parameter $\eta_p$. The cross section of the core of the waveguide was considered to be rectangular, as shown in figure 2, where the mode profile is also depicted, for two different heights of the waveguide core. The material of the core was chosen to be silicon nitride (SiN), which has the advantage that its refractive index is relatively insensitive to temperature fluctuations [16, 17]. It is supposed that the core is developed on a layer of $SiO_2$, as usually happens in practice. The environment was chosen to be water, which is the most common solvent, and it is also suitable for biological applications. The free space wavelength was supposed to be $\lambda_0 = 1550$ nm, which is a common wavelength emitted by fiber lasers, and further corresponds to the minimum of the fiber losses [18].

The refractive indices of the materials used in the simulations at the aforementioned wavelength are [19] $n_{SiN} = 2.463$, $n_{SiO_2} = 1.443$ and $n_{water} = 1.318$ for the silicon nitride, silicon dioxide and water respectively. The dimensions of the waveguide core were chosen to be $h = \lambda_{env}$ and $w = \lambda_{env}/4$, where $h$ is the height and $w$ the width of the waveguide, while $\lambda_{env} = \lambda_0 / n_{env}$ is the wavelength in the environment of the sensor, water in our case. This choice for the dimensions of the waveguide core ensures single mode operation and sufficient mode spreading. In order to optimize further the design of the refractive index sensor the parameter $\eta_p$ is calculated as function of the core width $w$ with the height fixed to $h = \lambda_{env}$. The results are shown in figure 3, where it is clear that the maximum value of $\eta_p$ is obtained for $w = 0.175\, \lambda_{env}$.

In order to optimize the height of the waveguide, the parameter $\eta_p$ is calculated as function of the waveguide height, with the width fixed to $w = 0.175\, \lambda_{env}$. The results are shown in figure 4. It is clear that the value of $\eta_p$ does not change much for $h \geq \lambda_{env}$. Further, for $h > 3\lambda_{env}/2$, multimode behavior is exhibited, which is not desirable for practical applications, as explained in the previous section. Thus, the optimum width of the waveguide core is $w_{opt} = 0.175\ \lambda_{env}$ while the optimum height lies in the region $\lambda_{env} \leq h_{opt} \leq 3\lambda_{env}/2$.

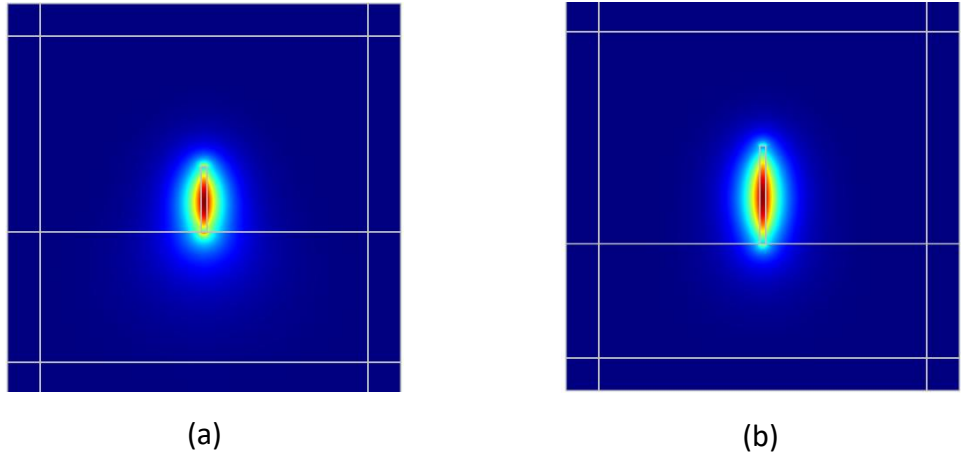

(a)            (b)

**Figure 2**: *The profile of the fundamental mode in a single core waveguide in the case of (a) $h = \lambda_{env}$ and (b) $h = 3\lambda_{env}/2$. In both cases the waveguide width is $w = 0.175\ \lambda_{env}$*

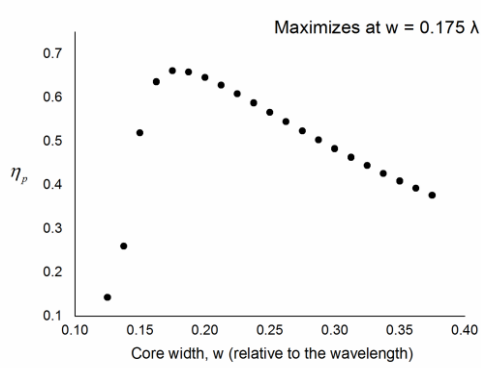 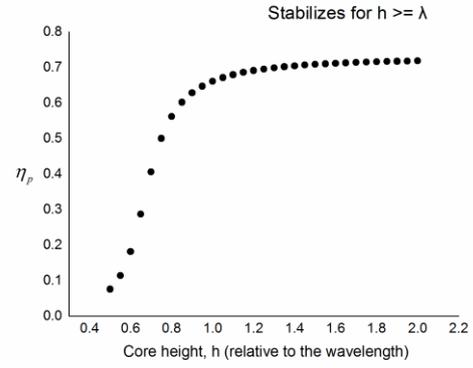

**Figure 3:** *The values of the parameter $\eta_p$ as function of the core width $w$ in the case of $h = \lambda_{env}$*

**Figure 4:** *The values of the parameter $\eta_p$ as function of the core height $h$ in the case of $w = 0.175\lambda_{env}$*

A double core configuration has also been studied, as shown in figure 5, where the mode profile is depicted. In this case the dimensions of each core were chosen to be $w = 0.175\ \lambda_{env}$ and $h = \lambda_{env}$. Higher values of $h$ have not been considered because the system exhibits multi-mode behavior. In order to find the optimum distance between the fiber cores, the parameter $\eta_p$ has been calculated as function of the core separation $d$. The results are shown in figure 6.

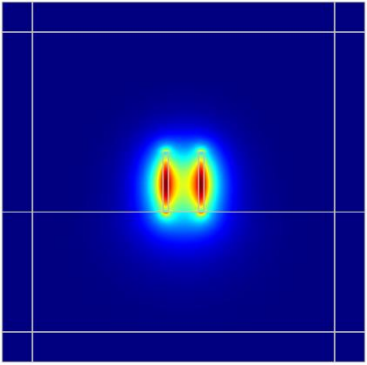

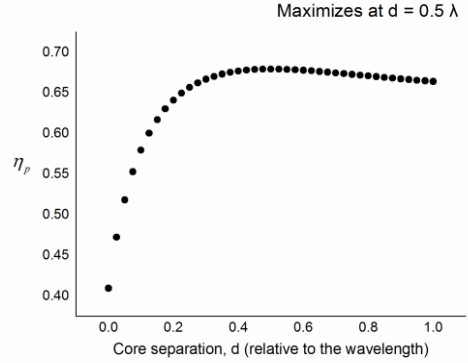

**Figure 5**: *The profile of the fundamental mode in a dual core waveguide in the case of $h = \lambda_{env}$, $w = 0.175\,\lambda_{env}$ and $d = \lambda_{env}/2$.*

**Figure 6:** *The values of the parameter $\eta_p$ as function of the core separation $d$ in the case of $w = 0.175\lambda_{env}$ and $h = \lambda_{env}$*

It is clear that $\eta_p$ is maximized for $d = \lambda_{env}/2$. Thus, the optimum conditions for a dual core design is $w_{opt} = 0.175\,\lambda_{env}$, $h_{opt} = \lambda_{env}$ and $d_{opt} = \lambda_{env}/2$. It should be noted that although these optimal values refer to an incident wavelength of 1550 nm, they can be used as guidelines for other wavelengths. This is due to the fact that the Maxwell equations are invariant under scale transformations and consequently the behavior of the system does not change if its dimensions are scaled according to the wavelength ratio. For example, if the wavelength is multiplied by a factor $x$ and all the dimensions of the waveguide are multiplied by the same factor, then its behavior will remain unchanged. Of course, in practice this cannot be achieved because the optical properties of the materials change due to dispersion. However, in any case, these values can be used as good starting points for design optimization. Next, the sensitivity is calculated for the three optimal configurations, specifically $w = 0.175\,\lambda_{env}$, $h = \lambda_{env}$, single core, $w = 0.175\,\lambda_{env}$, $h = 3\lambda_{env}/2$, single core, and $w = 0.175\,\lambda_{env}$, $h = \lambda_{env}$, $d = \lambda_{env}/2$, dual core. The results are shown in Table 1.

**Table 1**: *The basic characteristics of the three optimal designs described above.*

| No. of cores | $w/\lambda_{env}$ | $h/\lambda_{env}$ | $d/\lambda_{env}$ | $n_{eff}$ | $dn_{eff}/dn$ | $S$ |
|---|---|---|---|---|---|---|
| 1 | 0.175 | 1 | - | 1.46951 | 0.317 | 0.216 |
| 1 | 0.175 | 1.5 | - | 1.50435 | 0.330 | 0.219 |
| 2 | 0.175 | 1 | 0.5 | 1.50773 | 0.365 | 0.242 |

For these values of the sensitivity, equation (14) implies that the relative change in the resonant wavelength is $\Delta\lambda_{0,res}/\lambda_{0,res} = 0.216 \Delta n$ for a single core design $(h = \lambda_{env})$ and further increases by a factor of 1.014 in the case of $h = 3\lambda_{env}/2$ and 1.120 for a dual core configuration. This practically means that if the experimental setup is able to resolve a wavelength shift of $\Delta\lambda/\lambda \approx 3\times10^{-5}$, something that can be easily achieved by modern high-end optical spectrum analyzers (see for example [20]), then the system can sense a refractive index change of $\Delta n_{min} \simeq 1.4\times10^{-4}$ for a single core configuration $(h = \lambda_{env})$, which is further reduced by a factor of 1.014 in the case of $h = 3\lambda_{env}/2$ and 1.120 in the case of a dual core design.

It should be noted that the gain in the sensitivity for a dual core design is not that significant in order to justify the complexity of the design and the manufacturing challenges that entails. Therefore, I guess that in most practical situations, the design of choice would be a single core waveguide with dimensions $w = 0.175\,\lambda_{env}$ and $\lambda_{env} \leq h \leq 3\lambda_{env}/2$. Furthermore, as it will be shown in the following section, the choice $w = 0.175\,\lambda_{env}$ and $h = 3\lambda_{env}/2$ for the core dimensions provides better sensitivity than the dual core design as far as the behavior of the system as biosensor is considered.

### 4. Study of the bio-sensing properties of the system

In the following, the behavior of the system as biosensor is considered. For this purpose, a thin layer of biomaterial is supposed to be attached on the surface of the waveguide core. The dependence of the parameter $\eta_p$, namely the fraction of the mode power propagating within the layer, as function of the layer thickness is shown in figure 7. It is clear that $\eta_p$, and consequently the sensitivity, increases linearly for small values of the layer thickness, but exhibits some saturation as $t$ increases beyond $\sim 40\,nm$. It should also be noted that in this case the sensitivity of the single core waveguide with height $h = 3\lambda_{env}/2$ is greater than that of the dual core one.

In order to study the sensing ability of the system as the refractive index of the bio-layer changes, the relative change in the effective refractive index has been calculated as function of the effective index difference of the layer material, for a layer thickness of $t = 10$ and $t = 20\,nm$. The results are shown in figure 8. It is clear that the dependence is linear, and further the single core waveguide with height $h = 3\lambda_{env}/2$ exhibits the highest sensitivity, with a slope of 0.026 in the case of $t = 10\,nm$ and 0.050 when the layer thickness doubles.

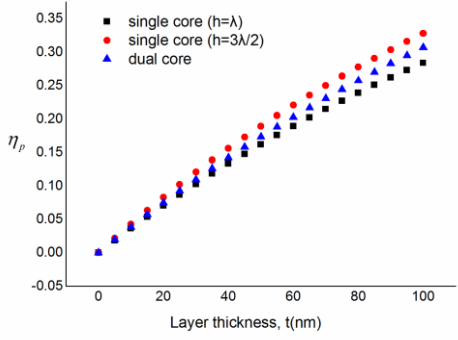 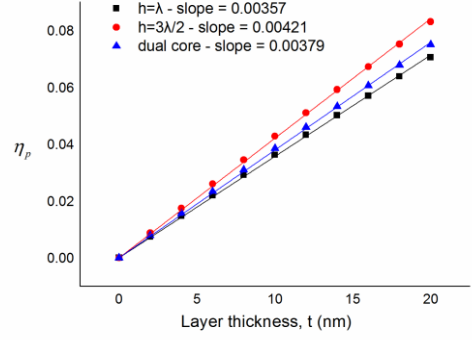

(a)　　　　　　　　　　　　　　　(b)

**Figure 7:** *The values of the parameter $\eta_p$ as function of the layer thickness $t$ for a range of (a) $100\ nm$ and (b) $20\ nm$ (zooming)*

This practically means that if the experimental setup can resolve a relative change in the resonant wavelength of the order of $3 \times 10^{-5}$, then the system can sense refractive index changes within the bio-layer of the order of $\Delta n_{\min} \simeq 1.15 \times 10^{-3}$ for a layer thickness of $10\ nm$, which is reduced by a factor of almost two when the layer thickness is doubled. Consequently, the system is able to detect very subtle changes in the composition or in the environment of the bio-layer.

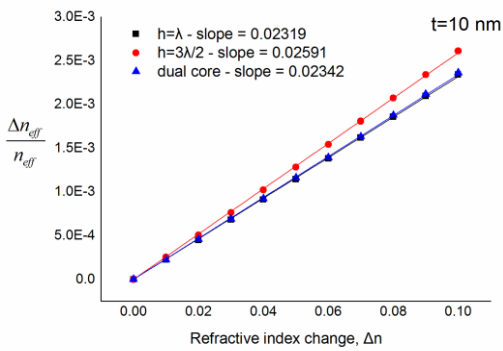 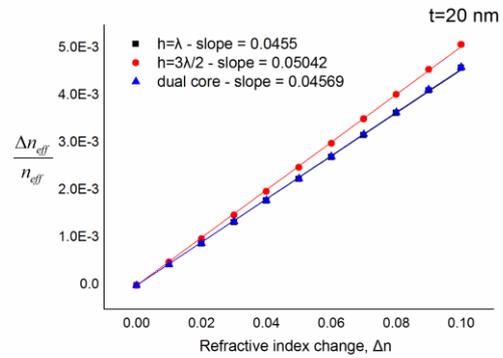

(a)　　　　　　　　　　　　　　　(b)

**Figure 8:** *The relative change in the effective refractive index, and consequently in the resonant wavelength, as function of the refractive index change $\Delta n$ in the bio-layer, when its thickness is (a) $t = 10\ nm$ and (b) $t = 20\ nm$*

Finally, the sensing ability of the system regarding the thickness of the bio-layer has been examined. For this purpose, the values of the relative change in the effective refractive index have been calculated as function of the layer thickness, supposing that the refractive index change within the layer is $\Delta n = 0.01$ and $\Delta n = 0.05$. The results are shown in figure 9.

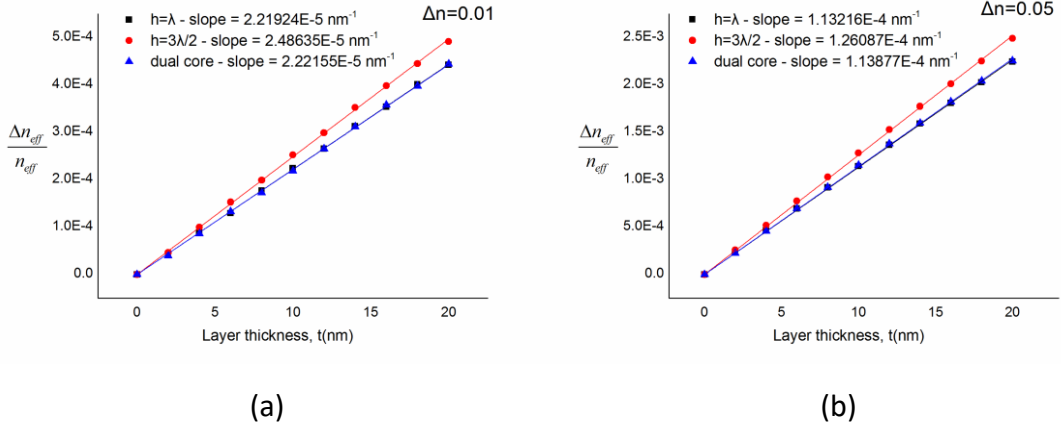

(a)                  (b)

**Figure 9:** *The relative change in the effective refractive index, and consequently in the resonant wavelength, as function of the thickness $t$ of the bio-layer, when the change in its refractive index is (a) $\Delta n = 0.01$ and (b) $\Delta n = 0.05$*

It is clear that the dependence is again linear, and the highest sensitivity is exhibited by the $h = 3\lambda_{env}/2$ single core waveguide. Indeed, the slope in this case is $2.49 \times 10^{-5} nm^{-1}$ for $\Delta n = 0.01$ and increases to $1.26 \times 10^{-4} nm^{-1}$ in the case of $\Delta n = 0.05$. Consequently, supposing that the minimum relative wavelength shift that can be measured is $3 \times 10^{-5}$, the minimum change in the layer thickness that can be detected is $\Delta t_{min} \simeq 1.2\ nm$ in the case of $\Delta n = 0.01$, which is further reduced to $\Delta t_{min} \simeq 0.24\ nm$ for $\Delta n = 0.05$. This impressive result leads to the conclusion that the system can detect even the slightest changes in the configuration of the biomolecules.

## 5. Conclusions

In this work the behavior of optical micro-ring resonators when functioning as refractive index sensors has been studied in detail. The sensitivity of the sensors has been defined and design guidelines in order to maximize its value are provided. It is found that, when optimally designed, the system can detect relative refractive index changes of the order of $\Delta n/n \approx 10^{-4}$, assuming that the experimental setup is able to resolve relative wavelength shifts of the order of $\Delta\lambda/\lambda \approx 3 \times 10^{-5}$. The performance of the systems as bio-sensor has also been examined. It is found that, when optimally designed, the system can detect refractive index changes of the order of $\Delta n \approx 10^{-3}$ for a layer thickness of $t = 10\ nm$, and changes in the layer thickness of the order of $\Delta t \approx 0.24$ n$m$, for a refractive index change of $\Delta n = 0.05$.